\documentclass{aastex631}

\newcommand{\hi}{\mbox{H\,{\sc i}}}

\newcommand{\corrections}[1]{\textcolor{blue}{#1}}

\begin{document}

\title{\emph{Fermi} Unassociated Sources in the MeerKAT Absorption Line Survey}

\author[0009-0004-1469-9550]{Morgan Himes}
\affiliation{National Radio Astronomy Observatory, 1003 Lopezville Road, Socorro, NM 87801, USA}
\affiliation{University of Florida, Gainesville, FL 32611, USA}

\author[0000-0002-5825-9635]{Preshanth Jagannathan}
\affiliation{National Radio Astronomy Observatory, 1003 Lopezville Road, Socorro, NM 87801, USA}

\author[0009-0008-7604-003X]{Dale A. Frail}
\affiliation{National Radio Astronomy Observatory, 1003 Lopezville Road, Socorro, NM 87801, USA}

\author[0000-0001-6672-128X]{Frank Schinzel}
\affiliation{National Radio Astronomy Observatory, 1003 Lopezville Road, Socorro, NM 87801, USA}

\author[0000-0001-7547-4241]{Neeraj Gupta}
\affiliation{Inter University Centre for Astronomy and Astrophysics, Pune, Maharashtra 411007, India}

\author[0000-0002-3814-9666]{S. A. Balashev}
\affil{Independent Researcher}

\author[0000-0003-2658-7893]{F. Combes}  
\affil{ Observatoire de Paris, Coll\`ege de France, PSL University, Sorbonne University, CNRS, LERMA, Paris, France}

\author[0000-0002-9931-1313]{P. P. Deka}
\affil{Inter-University Centre for Astronomy and Astrophysics, Post Bag 4, Ganeshkhind, Pune 411 007, India}

\author[0000-0002-0648-2704]{H.-R. Kl\"ockner}  
\affil{Max-Planck-Institut f\"ur Radioastronomie, Auf dem H\"ugel 69, D-53121 Bonn, Germany}

\author[0000-0003-3168-5922]{E. Momjian}  
\affil{National Radio Astronomy Observatory, Socorro, NM 87801, USA}

\author[0000-0003-1321-0886]{J. D. Wagenveld}  
\affil{Max-Planck-Institut f\"ur Radioastronomie, Auf dem H\"ugel 69, D-53121 Bonn, Germany}



\begin{abstract}

Over 2000 $\gamma$-ray sources identified by the Large Area Telescope (LAT) on NASA's \emph{Fermi} Gamma-ray Space Telescope are considered unassociated, meaning that they have no known counterparts in any other frequency regime. We have carried out an image-based search for steep spectrum radio sources, with in-band spectral index $<-1.4$, within the error regions of \emph{Fermi} unassociated sources using 1-1.4 GHz radio data from the MeerKAT Absorption Line Survey (MALS) Data Release.  MALS DR1 with a median rms noise of 22-25\,$\mu$Jy and 735,649 sources is a significant advance over past image-based searches with improvements in sensitivity, resolution and bandwidth. Steep spectrum candidates were identified using a combination of in-band spectral indices from MALS and existing radio surveys. We developed an optical and infrared source classification scheme in order to distinguish between galactic pulsars and radio galaxies. In total, we identify nine pulsar candidates towards six \emph{Fermi} sources that are worthy of follow-up for pulsation searches. We also report 41 steep spectrum radio galaxy candidates that may be of interest in searches for high-redshift radio galaxies.
We show that MALS due to its excellent continuum sensitivity can detect 80\% of the known pulsar population. This exhibits the promise of identifying exotic pulsar candidates with future image-based surveys with the Square Kilometre and its precursors.

\end{abstract}



\section{Introduction} \label{sec:intro}

The most recent data release of the Large Area Telescope \citep{atwood2009large}, a part of NASA's \emph{Fermi} Gamma-ray Space Telescope, provides 14 years of survey data of the entire sky in the high energy $\gamma$-ray regime. Of the 7195 sources in the LAT 14-year Source Catalog (4FGL-DR4; see \cite{ballet2023fermi}), 2427 are considered unassociated sources (UAS). These sources have no known counterparts in any other frequency regime. Many sources of $\gamma$-ray emission, such as active galactic nuclei (AGN) or pulsars, detected by \emph{Fermi} would be expected to appear in radio observations \citep{2021ApJ...914...42B}. 

Pulsation searches, both at radio and gamma-ray wavelengths, have proven very effective at identifying new millisecond pulsars (MSP) and young, energetic pulsars \citep{smith2023fermi}.  
Increasingly, image-based methods are being used at X-ray, optical and radio wavelengths to identify promising \emph{Fermi} pulsar candidates \citep[e.g.,][]{2017ApJ...844..150H,2013ApJ...779..133A,2021ApJS..252...17B}. In radio images the tell-tale signatures that distinguish a pulsar from the background of (mostly) extragalactic sources are compactness, steep spectrum, polarization and variability. The first of these two criteria have been used with modest success yielding detections of \emph{Fermi} MSPs \citep{2018MNRAS.475..942F,2022ApJ...927..216R}. Looking for high fractional polarization shows promise because of its lower false-positive rate, but it has not yet been used systematically on UAS \citep[but see][]{2022A&A...661A..87S}. Short-term variability in the order of seconds, induced by interstellar scintillation \citep{2016MNRAS.462.3115D}, fell out of favor after being used to identify the first MSP \citep{1982Natur.300..615B}, but it is experiencing a recent resurgence \citep{2022MNRAS.516.5972W,2023PASA...40....3S,2023MNRAS.525L..76H}

One of the more efficient image-based techniques for finding pulsar candidates among the \emph{Fermi} UAS is through cross-matching with {\it existing} radio sky surveys. This method has been previously used to identify pulsar candidates from large surveys such as the Giant Metrewave Radio Telescope (GMRT) 150 MHz All-Sky Radio Survey \citep{frail2016pulsar}.  The various shortcomings of this image-based methods have been discussed by \cite{2018MNRAS.475..942F} but these include low angular resolution (~10"), limited fractional bandwidths ($<$10\%) and the low sensitivity of existing synoptic radio surveys \citep{2017ApJ...838..139S,2023ApJ...943...51B}. Fortunately, a new generation of decameter to centimeter radio surveys are underway that represents a substantial improvement in capabilities suited for deeper searches of pulsar candidates \citep{2022PASA...39...35H,2023MNRAS.523.1729B,2023PASA...40...34D,2023A&A...673A.165D}.

In this paper we use the newly acquired data from the southern hemisphere MeerKAT Radio Telescope as part of the MeerKAT Absorption Line Survey \citep[MALS;][]{gupta2017meerkat} to identify new pulsar candidates within \emph{Fermi} unassociated 95\% confidence error ellipses. In \S{2} we describe the properties of MALS and our search methodology. In \S{3} we describe our final candidates and we end in \S{4} with a description of the limitations of image-based search methods and the prospects for improvements.

\section{Methodology} \label{sec:method}
\subsection{The MeerKAT Absorption Line Survey (MALS)}

The MeerKAT Absorption Line Survey (MALS) is a continuum and spectral line survey primarily designed to carry out a search of intervening and associated neutral hydrogen (\hi) 21-cm and the hydroxyl radical (OH) 18-cm absorption lines at 0$<z<$2 unbiased by dust obscuration \citep[][]{gupta2017meerkat}. Each of the approximately 500 pointings \corrections{($1.1^{\circ} < b < 84^{\circ} $)}, with either the L-band (900 MHz-1670 MHz) or/and UHF-band (580-1015 MHz), are centered on a radio source i.e., AGN brighter than 200 mJy at 1 GHz. However, the field-of-view of the MeerKAT telescope is sufficiently large \citep[88$^\prime$ full width at half maximum at 1 GHz;][]{2016mks..confE...1J} that a variety of commensal science can be carried out. 

For this work we used MALS Stokes $I$ image catalogs from the first data release \citep[DR1;][]{deka2023meerkat}. The DR1 is based on 391 telescope pointings at $\delta\lesssim+20^{\circ}$ observed using MeerKAT's L-band (900-1670 MHz) during 2020, April, 01 to 2021, January, 18. The release consists of flux density measurements for 15 spectral windows covering L-band \citep[][their Fig.~2]{Gupta21}. It also provides in-band spectral indices calculated using the measurements from the spectral windows 2 (960-1010 MHz) and 9 (1450-1500 MHz), with central frequencies of 1006.0 MHz and 1380.9 MHz, respectively.  The survey achieves a median rms noise of 25 $\mu$Jy beam$^{-1}$ for 1006.0 MHz and 22 $\mu$Jy beam$^{-1}$ for 1380.9 MHz with spatial resolution of $12^{\prime\prime}$ and $8^{\prime\prime}$,  respectively. The survey detects 495,325 and 240,321 sources at a signal-to-noise ratio (SNR) $>$ 5 over an area of 2289 deg$^2$ at 1006.0 MHz and 1132 deg$^2$ at 1380.9 MHz.  The astrometric accuracy is estimated to be better than $1^{\prime\prime}$, and the flux density accuracy is estimated at 6 percent.  

\subsection{Known Pulsars in MALS}

As an initial check on the sensitivity of MALS we looked for {\it known} pulsars in the ATNF Pulsar Catalogue \citep{manchester2005australia}. There are 56 known pulsars that lie within the MALS pointings, of which there are 1.4 GHz mean flux densities published for only 26. 24 of the pulsars should be detectable with MALS sensitivity at SNR $>$ 5 but we detect  only 14, including two pulsars without published flux densities. Thus, the efficiency of MALS detection rate is slightly in excess of 50\%, a fraction that is comparable to other recent image-based searches \citep[e.g.,][]{frail2016pulsar,2023PASA...40....3S}. Short-term variability of the pulsar, occurring on a timescale comparable to the MALS integration time of 56\,minutes per pointing, likely explains why the 12 remaining bright pulsars were not detected.

Three of the 14 MALS pulsars (J1221-0633, J2124-3358, and J2256-1024) are known millisecond pulsars in the Third Fermi Large Area Telescope Catalog of Gamma-ray Pulsars \citep{smith2023fermi}. The other 11 pulsars are not known to be associated with any gamma-ray sources and do not lie within any of the \emph{Fermi} error ellipses. Table \ref{Tab:psrs} summarizes the properties of the known pulsars detected in MALS.

\begin{deluxetable}{llllcccccc}
\caption{A summary of the 14 known pulsars detected in MALS. Three of these pulsars (J1221-0633, J2124-3358, and J2256-1024) are known millisecond pulsars in the Third Fermi Large Area Telescope Catalog of Gamma-ray Pulsars \citep{smith2023fermi}. The dispersion measure (DM), 1.4 GHz flux density (S1.4) and barycentric period (P) of the pulsar are from the ATNF Pulsar Catalogue .}
\tablehead{\colhead{PSR Name} & \colhead{MALS Source} & \colhead{RA (h m s)} & \colhead{DEC ($^\circ$\ $'$\ $"$)}& \colhead{$\alpha$} & \colhead{DM (pc cm$^{-3}$)} & \colhead{P (s)} & \colhead{S1.4 (mJy)}}
\startdata
B0148-06 & J015122.71-063502.9 & 01 51 22.71  ($\pm$ 0.13) & -06 35 02.94  ($\pm$ 0.15) & -2.97 \tablenotemark{a} & 25.66 & 1.4647 & 1.6 \\
J0520-2553 & J052036.19-255312.6 & 05 20 36.20 ($\pm$ 0.25) & -25 53 12.69 ($\pm$ 0.35) & -0.66 \tablenotemark{a} & 33.77 & 0.2416 & 0.8 \\
J0900-3144 & J090043.94-314430.5 & 09 0 43.94  ($\pm$ 0.03) & -31 44 30.50  ($\pm$ 0.04) & -1.77 \tablenotemark{a} & 75.69 & 0.0111 & 3.84 \\
J1221-0633 & J122124.72-063351.7 & 12 21 24.73  ($\pm$ 0.43) & -06 33 51.73  ($\pm$ 0.58) & -4.84 \tablenotemark{b} & 16.43 & 0.0019 & \\
B1237-41 & J124017.50-412451.1 & 12 40 17.51 ($\pm$ 0.55) & -41 24 51.17 ($\pm$ 0.65) & -1.6 \tablenotemark{a} & 44.1 & 0.5122 & 0.6\\
B1325-49 & J132833.45-492134.2 & 13 28 33.45 ($\pm$ 0.06) & -49 21 34.29 ($\pm$ 0.09) & -3.16 \tablenotemark{c} & 118.0 & 1.4787 & 0.82\\
J1346-4918 & J134622.42-491806.3 & 13 46 22.43 ($\pm$ 0.84) & -49 18 06.32 ($\pm$ 1.32) & -6.21 \tablenotemark{a} & 74.42 & 0.2996 & 0.5\\
B1620-09 & J162317.67-090848.5 & 16 23 17.67 ($\pm$ 0.49) & -09 08 48.57 ($\pm$ 1.16) & -1.6 \tablenotemark{d} & 68.18 & 1.2764 & 0.6\\
B1726-00 & J172834.75-000744.0 & 17 28 34.75 ($\pm$ 0.33) & -00 07 44.03 ($\pm$ 0.59) & -1.21 \tablenotemark{a} & 41.09 & 0.386 & 0.66\\
J1738+0333 & J173853.93+033311.5 & 17 38 53.93 ($\pm$ 0.32) & +03 33 11.5 ($\pm$ 0.5) & -3.86 \tablenotemark{a} & 33.77 & 0.0059 & 0.34\\
J1932-3655 & J193206.06-365501.1 & 19 32 6.06 ($\pm$ 0.27) & -36 55 01.11 ($\pm$ 0.26) & -2.65 \tablenotemark{a} & 59.88 & 0.5714 & 0.7\\
J2124-3358 & J212443.80-335845.9 & 21 24 43.81 ($\pm$ 0.22) & -33 58 45.90 ($\pm$ 0.11) & -1.31 \tablenotemark{e} & 4.6 & 0.0049 & 4.5\\
B2122+13 & J212446.64+140719.2 & 21 24 46.65($\pm$ 1.52) & +14 07 19.26 ($\pm$ 2.25) &  & 30.25 & 0.6941 & \\
J2256-1024 & J225656.33-102434.8 & 22 56 56.33 ($\pm$ 0.25) & -10 24 34.80 ($\pm$ 0.34) & 0.02 \tablenotemark{a} & 13.78 & 0.0023 & 0.73
\enddata
\tablenotetext{a}{Calculated from MALS 1006.0 MHz and ATNF 1400 MHz flux densities \citep{deka2023meerkat, manchester2005australia}.}
\tablenotetext{b}{Calculated from MALS 1006.0 MHz and 3rd Fermi LAT Pulsar Catalog 1400 MHz flux densities \citep{deka2023meerkat, smith2023fermi}.}
\tablenotetext{c}{Calculated from RACS-low1 887.5 MHz, MALS 1006.0 MHz, and MALS 1380.9 MHz flux densities \citep{mcconnell2020rapid, hale2021rapid, deka2023meerkat}.}
\tablenotetext{d}{Retrieved from TGSS-ADR1 \citep{de2018radio}.}
\tablenotetext{e}{Calculated from RACS-low1 887.5 MHz, MALS 1006.0 MHz, and ATNF 1400 MHz flux densities \citep{mcconnell2020rapid, hale2021rapid, deka2023meerkat, manchester2005australia}.}
\label{Tab:psrs}
\end{deluxetable}

\subsection{Matching 4FGL Unassociated Sources}


The MALS catalog for spectral windows 2 and 9, containing 735,649 sources, was crossmatched with the  unassociated sources in the \emph{Fermi} 4FGL-DR4 \citep{2022ApJS..260...53A}. The MALS catalog was first filtered to include only compact sources with reported spectral indices. The compactness as defined in the catalog contains sources which are unresolved and where the source flux is well represented by a peak to total flux density ratio of less than 1.2. 
A total of 657,318 are compact and 551,069 of these compact sources have in-band spectral indices. 

Sources with ultrasteep negative spectral indices (positive convention, $S_{\nu}\propto\nu^{+\alpha}$) are targeted because pulsars typically exhibit spectral indices $\alpha<-1.4$ \citep{bates2013pulsar}. Ultrasteep spectrum radio point sources within \emph{Fermi} error ellipses could be millisecond pulsars. If follow-up radio interferometric observations at higher resolutions show resolved structure, the sources are instead likely to be AGN or high-redshift radio galaxies (HzRGs). 

These compact sources were cross-matched with the unassociated \emph{Fermi} error ellipses. There are 1,090 compact MALS sources located within 74 Fermi error ellipses. Of these, 883 sources within 41 ellipses have spectral indices. Then, we filtered out sources that had spectral index error flags in the catalog (i.e., $\pm$999), resulting in a candidate set of 604 sources within 38 ellipses. Finally, we filtered this set to sources with spectral indices $\alpha<-1.4$, resulting in 95 sources towards 14 error ellipses. Because the MALS catalog reports detections in both spectral windows 2 and 9 as separate detections, these sources were matched to each other and combined to be the same object. Not all the 95 sources were detected in both bands and once duplicates were removed we were left with 50 unique compact steep spectrum sources within the \emph{Fermi} error ellipses.

\subsection{Source Classification Scheme}

Next we adopted a classification scheme for our 50 candidates in an effort to distinguish between galactic pulsars and extragalactic radio sources (e.g., blazar or AGN). For all candidates we used the CIRADA cutout service to obtain radio images from the VLA All-Sky Survey \citep[VLASS-QL;][]{2020PASP..132c5001L}, the TIFR GMRT Sky Survey \citep[TGSS ADR1;][]{2017A&A...598A..78I}
and the Rapid ASKAP Continuum Survey \citep[RACS-low1;][]{mcconnell2020rapid, hale2021rapid}, mid-infrared images from the Wide-field Infrared Survey Explorer \citep[WISE;][]{wright2010wide, mainzer2011preliminary}, and optical/NIR Pan-STARRS1 images \citep{chambers2016pan} when available. We also acquired images from the Dark Energy Camera Legacy Survey where available \citep[DECaLS;][]{dey2019overview}. 

The purpose of the additional radio images  was to further constrain the compactness and/or spectral index of the candidates (from 150 MHz to 3 GHz). Optical and infrared images were used to construct color-color plots. Most of the steep spectrum MALS sources with WISE color matches fell into known AGN planes on a WISE color-color diagram
\citep{mateos2012using, chang20172whsp, gupta2022mals}. Those that do not (or do not have WISE color matches) are deemed best for pulsar candidate searches.

\begin{figure}[htb!]
\includegraphics[width=11cm]{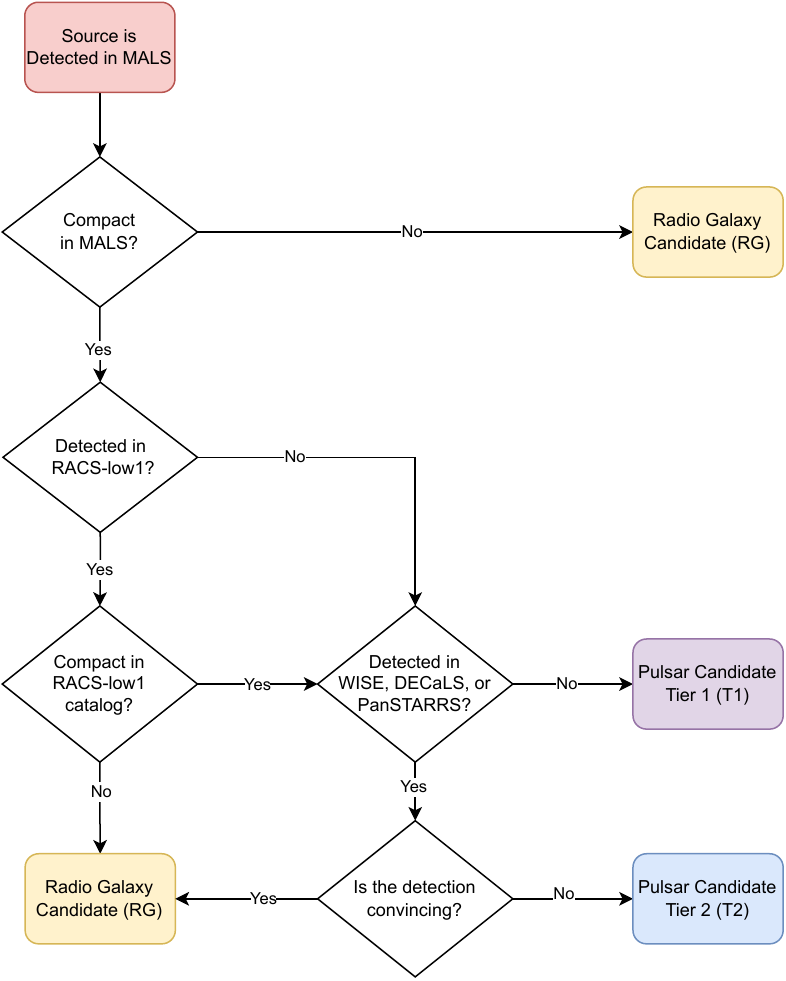}
\centering
\caption{The classification scheme used for the 50 steep spectrum MALS sources within Fermi error ellipses. Using WISE, PanSTARRS, DECaLS, and RACS data, the sources are sorted into three categories: Tier 1 (T1) and Tier 2 (T2) pulsar candidates, and radio galaxy (RG) candidates. The sources and candidate types can be found in Table \ref{tab:allsrc}. To determine if a detection in WISE, DECaLS, or PanSTARRS is convincing, we inspect the source image for any offsets from the source position or evidence that the detection is an imaging artifact. \label{fig:flowchart}}
\end{figure}

We followed a general classification framework to organize sources into three categories: Radio Galaxy (RG), Pulsar Tier 1 (T1), and Pulsar Tier 2 (T2). Pulsar Tier 1 candidates are defined as the most likely to be pulsars and should be prioritized for follow-up higher resolution observations. Pulsar Tier 2 candidates are sources we were unable to rule out from pulsar consideration but suggest possible evidence of being galaxies. See Figure \ref{fig:flowchart} for a flowchart of the classification process.

With these criteria, the 50 sources were divided into 41 RG candidates, 5 T2 candidates, and 4 T1 pulsar candidates. The sources and candidate types can be found in Table \ref{tab:allsrc}.

\section{Pulsar Candidates}

Tier 1 pulsar candidates are selected based on their steep spectral indices, compact appearance in MALS images, and non-detection in other surveys (VLASS, RACS, PanSTARRS, WISE, DECaLS). The source properties are listed in Table~\ref{tab:tier1}. It is worth noting that the sources $4FGL J0041.3-0048$ and $4FGL J1334.8-3856$ have listed possible associations (probability 0.51 and 0.52) as blazars. 

\begin{table}[h]
\centering
\begin{tabular}{c c c c c}
\hline
Fermi Source& MALS Source & Distance (') & S1.4(mJy) & $\alpha$ \\
\hline
4FGL J0041.3-0048 & J004151.00-010950.5 & 23.06 arcmin & 1.89 $\pm$ 0.49 mJy & $-1.56$ \\ 
4FGL J0634.6-3046c & J063429.94-304642.1 & 2.22 arcmin & 0.51 $\pm$ 0.10 mJy & $-1.46$ \\ 
4FGL J1245.4-0701 & J124518.45-070133.3 & 1.48 arcmin & 0.65 $\pm$ 0.29 mJy & $-1.77$ \\ 
4FGL J1334.8-3856 & J133504.13-385859.0 & 3.94 arcmin & 0.25 $\pm$ 0.08 mJy & $-1.60$ \\ 
\hline
\end{tabular}
\caption{Tier 1 Pulsar Candidates}
\label{tab:tier1}
\end{table}

Tier 2 Candidates are listed in Table.~\ref{tab:tier2}. These pulsars have detection in other surveys besides MALS. Furthermore, the source $4FGL J2052.3-2707$ has been classified as possible blazar in DR4 with association probability of (0.73).

\begin{table}[h]
\centering
\begin{tabular}{ c c c c c c }
\hline
Fermi Source& MALS Source & Distance (') & S1.4(mJy) & $\alpha$ & Notes \\
\hline
4FGL J1959.8-1835 & J195939.23-183924.2 & 5.21 arcmin & 1.05 $\pm$ 0.31  & $-1.73$ & Possible Wise Detection\\ 
4FGL J0252.3-2707 & J025300.17-270837.0 & 8.92 arcmin & 5.35 $\pm$ 0.77  & $-1.98$ & RACS-low: 11.86 +/- 3.35 mJy \\ 
\hline
\end{tabular}
\caption{Tier 2 Pulsar Candidates}
\label{tab:tier2}
\end{table}

See Table~\ref{tab:allsrc} for a summary of the other steep spectrum sources inside the error ellipse of this \emph{Fermi} source. We prioritize Tier 1 pulsar candidates as the most probable gamma-ray associations.

\section{Galaxy Candidates} \label{sec:galcand}

Sources that are not Tier 1 or Tier 2 pulsar candidates are classified as radio galaxy candidates. These sources appear either extended in MALS wideband images or have strong detections in WISE, PanSTARRS, or DECaLS. The properties of these sources are summarized in Table \ref{tab:allsrc}. Since the sources have steep radio spectra, it is possible that they may be high-redshift radio galaxy (HzRG) candidates and not primarily a \emph{Fermi} UAS.

A candidate of particular interest is MALS J162101.76+002317.4. Possible counterpart sources are detected in PanSTARRS and DECaLS and exhibit a galaxy-like structure. The photometric redshift of the source from the DESI Legacy Imaging Surveys is 0.451 \citep{duncan2022all}. The source was observed with the VLA on June 27, 2023 in BnA configuration at S-band (2-4 GHz) and L-band (1-2 GHz) for 13 minutes on-source time in each band. The source was not detected, possibly suggesting variability in the flux density. With further observations, it may be of interest as a variable source. 

\section{Discussion \& Conclusions} \label{sec:disc}

We have used the first data release of the MeerKAT Absorption Line Survey (MALS) in the L-band (1-1.4 GHz) to identify 50 compact ($\theta\lessapprox 10^{\prime\prime}$), steep spectrum ($\alpha<-1.4$) radio sources in coincidence with \emph{Fermi} unassociated gamma-ray sources. By crossmatching with other radio and infrared surveys, we sort these 50 sources into three categories: Tier 1 (T1) pulsar candidates, Tier 2 (T2) pulsar candidates, and radio galaxy (RG) candidates. T1 are considered to be our most likely pulsar candidates because they have no associations in any other survey that was crossmatched. We suggest the four T1 sources as higher priority for follow-up observations. 

\begin{figure}[htb!]
\includegraphics[width=17cm]{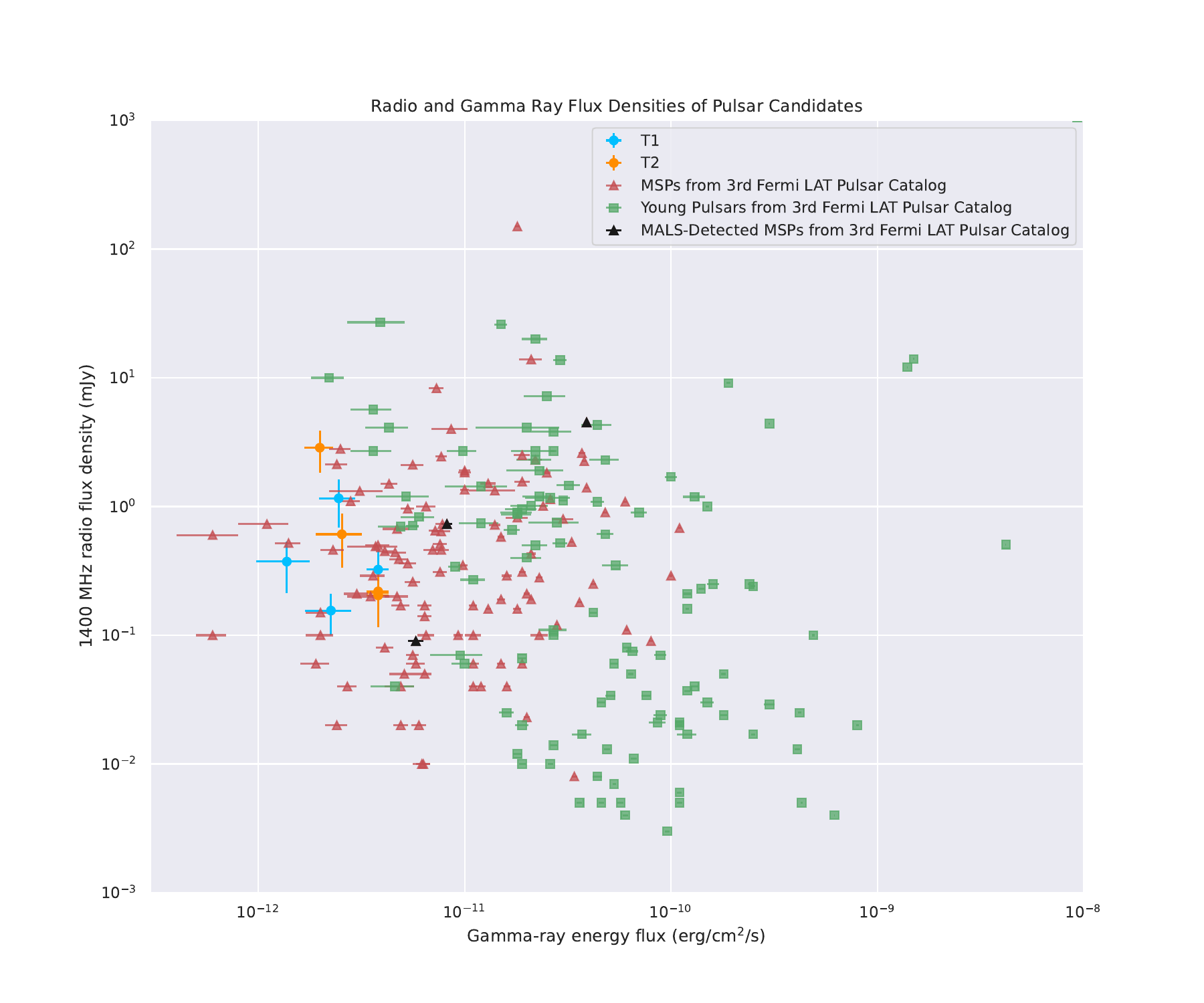}
\centering
\caption{Each Tier 1 (T1, blue circle) and Tier 2 (T2, orange circle) pulsar candidate's radio flux density is plotted against the gamma ray energy flux of the potential associated Fermi source. A selection of millisecond pulsars (red triangle) and young pulsars (green square) from the Third Fermi Large Area Telescope Catalog of Gamma-ray Pulsars are plotted \citep{smith2023fermi}. Three millisecond pulsars in the Third Fermi LAT Pulsar Catalog are detected in MALS and are plotted (black triangle). \label{fig:radiogammapsrs}}
\end{figure}

The radio flux densities  of T1 and T2 candidates are plotted against the gamma-ray fluxes of their \emph{Fermi} sources in Figure \ref{fig:radiogammapsrs}. These are plotted in addition to known millisecond pulsars and young pulsars from the Third Fermi Large Area Telescope Catalog of Gamma-ray Pulsars \citep{smith2023fermi}. The T1 and T2 data points appear to mostly fall into the region of the plot where many millisecond pulsars are located. As such, these particular candidates may be suitable targets for millisecond pulsar searches.

\begin{figure}[htb!]
\includegraphics[width=15cm]{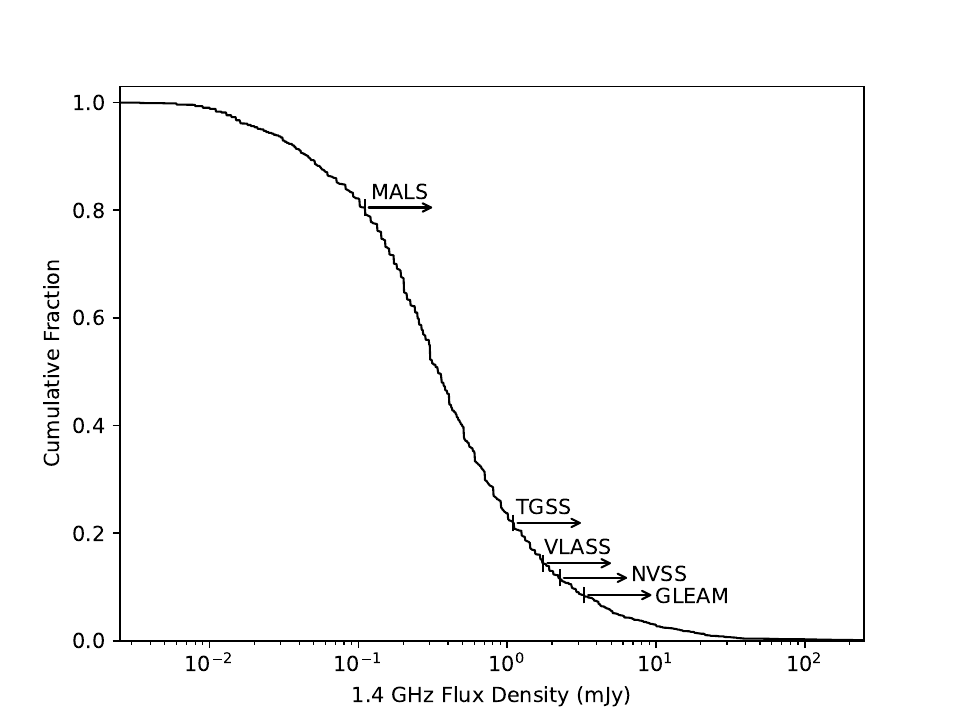}
\centering
\caption{Cumulative histogram of the 2374 known pulsars with published 1.4 GHz flux densities. We compare the capabilities of MALS to several radio surveys which have been used for image-based searches of pulsars. to do so we estimated the 5$\sigma$ rms noise for each survey and then scaled them to 1.4 GHz using our steep spectral index cutoff of $\alpha=-1.4$. Past surveys have been less sensitive to the bulk of the known pulsar population compared to the new generation of surveys being carried out on MWA, ASKAP and MeerKAT.\label{fig:surveyhistogram}}
\end{figure}

MALS is one of the new generation of surveys. Its improvements to sensitivity, frequency range, and angular resolution have distinct advantages for identifying pulsar candidates over past surveys such as NVSS, SUMSS, GLEAM and TGSS \citep{1998AJ....115.1693C,2017A&A...598A..78I,2017MNRAS.464.1146H}. We illustrate the sensitivity advantage of MALS in Figure \ref{fig:surveyhistogram} where we have plotted a cumulative histogram of the 2374 known pulsars with published 1.4 GHz flux densities. For ease of comparison, we have calculated the 5$\sigma$ rms noise for several decameter and centimeter radio surveys, scaled to 1.4 GHz using our steep spectral index cutoff of $\alpha=-1.4$. Most previous continuum surveys have been sensitive to between 7\% and 21\% of the {\it known} pulsar population. This is also true of the on-going VLA Sky Survey \citep[VLASS;][]{2020PASP..132c5001L} despite improved sensitivity, because of its relatively high observing frequency (2-4 GHz). MALS in contrast is sensitive to 80\% of the known pulsar population. Such sensitivity gains are important since in the absence of substantial interstellar scattering, pulsation searches always have an approximately 4x advantage over imaging \citep[i.e., ${\sqrt{1/f}}$, where f is the pulsar duty cycle][]{1985ApJ...294L..25D}. While not predictive, Figure~\ref{fig:surveyhistogram} shows the promise of future image-based surveys with MeerKAT, the Murchison Widefield Array (MWA) and Australian Square Kilometre Array Pathfinder (ASKAP).

The ability to measure in-band spectral indices is another advantage of this next generation of surveys. Past image-based pulsar searches used spectral indices calculated from two or more surveys taken at different frequencies and different telescopes, using data taken years or even decades apart \citep{2000A&AS..143..303D,de2018radio}. On these timescales AGN variability (and perhaps transients) could be a significant contaminant of the steep spectrum tail of the spectral index distribution \citep{2018MNRAS.475..942F}. A spectral index, measured across the observing bands, eliminates false positives that originate due to variability between epochs.

An ongoing challenge to identify pulsar candidates in total intensity images is the large number of background AGN. The MALS source density at 1380.9 MHz is approximately 210 sources/deg${^2}$ and the vast majority of these are AGN or star-forming radio galaxies. We were able to reduce the number of AGN substantially by using the MALS's excellent angular resolution and its measured in-band spectral indices. These compactness and spectral index criteria reduced the number of candidates within \emph{Fermi} UAS down to a total of 50; still a large number. By using optical/NIR color we were further able to show that 82\% of candidates in Table \ref{tab:allsrc} fell into well known AGN planes on a color-color diagram. To our knowledge, this is the first time such plots have been used to eliminate false positives in radio image-based searches for pulsars. 

Polarization, both linear and circular, is increasingly being used for identifying pulsars and candidates in the image plane either on its own or in combination with other criteria. \cite{1995ApJ...455L..55N} were the first to use polarization, compactness and a steep spectrum to find the luminous 2.3 msec pulsar PSR\,J0218+4232. While more recently, ASKAP has been finding pulsars this way \citep{2022ApJ...930...38W}, and the VLA has identified some promising candidates \citep{2021MNRAS.507.3888H}. The lower number of background AGN and the negligible confusion from diffuse Galactic emission compared to total intensity images have led to a number of recent surveys that use polarization to identify pulsar candidates \citep[e.g.,][]{2022A&A...661A..87S,2023PASA...40....3S}. 

In the future, we plan to use the Stokes V images from the MALS project, together with other criteria to identify promising pulsar candidates. Future MALS data releases will include wide-band and full Stokes images for each targeted field at both L- and UHF-bands. 
For example, the upcoming second MALS data release based on full wideband images corresponding to 391 MALS L-band pointings has image rms noise as low as 10\,$\mu$Jy beam$^{-1}$ near the pointing centres, and  consists of 971,980 sources detected over an area of 4344 deg$^2$ \citep[][]{Wagenveld24}.

Cross matching of \emph{Fermi} UAS using this expanded MALS dataset should identify more candidates in the southern hemisphere. Finally, we note that the remaining 41 sources are considered radio galaxy candidates; these are reported in Table \ref{tab:allsrc} as sources of interest for high-redshift radio galaxy (HzRGs) searches. An extension of this project will include analysis of all compact steep spectrum sources in MALS with particular emphasis on searching for HzRGs \citep{2018MNRAS.475.5041S}

\begin{deluxetable}{ccccccccc}
\tabletypesize{\footnotesize}
\caption{Data is reported for each of the 50 steep spectrum MALS sources that were identified within \emph{Fermi} error ellipses. The RA and DEC of the sources in MALS are given with errors in arcsec. The D column contains the distance between the MALS source and the Fermi source in arcmin. Each candidate is given a source type: T1 for Tier 1 pulsar candidate, T2 for Tier 2 pulsar candidate, and RG for radio galaxy. When available, photometric redshift values ("zphot" column) for radio galaxy candidates are acquired from the DESI Legacy Imaging Surveys \citep{duncan2022all}.}\label{tab:allsrc}
\tablehead{\colhead{MALS Source} & \colhead{Fermi Source} & \colhead{RA} & \colhead{DEC} & \colhead{D} & \colhead{SPIDX} & \colhead{zphot} & \colhead{Type}}
\startdata
J004151.00-010950.5 & 4FGL J0041.3-0048 & 0h41m51.00804s ($\pm 0.37$) & -1d09m50.5288224s ($\pm 0.49$) & 23.06 & -1.56 &  & T1 \\
J063429.94-304642.1 & 4FGL J0634.6-3046c & 6h34m29.93981s ($\pm 0.38$) & -30d46m42.157416s ($\pm 0.27$) & 2.22 & -1.46 &  & T1 \\
J124518.45-070133.3 & 4FGL J1245.4-0701 & 12h45m18.4517s ($\pm 1.65$) & -7d01m33.3774192s ($\pm 1.63$) & 1.48 & -1.77 &  & T1 \\
J133504.13-385859.0 & 4FGL J1334.8-3856 & 13h35m4.13002s ($\pm 0.46$) & -38d58m59.0124s ($\pm 0.76$) & 3.94 & -1.6 &  & T1 \\
J025300.17-270837.0 & 4FGL J0252.3-2707 & 2h53m0.17116s ($\pm 0.20$) & -27d08m37.069368s ($\pm 0.24$) & 8.92 & -1.98 &  & T2 \\
J063429.91-305411.0 & 4FGL J0634.6-3046c & 6h34m29.91172s ($\pm 0.37$) & -30d54m11.008296s ($\pm 0.31$) & 8.23 & -1.74 &  & T2 \\
J063520.46-304736.2 & 4FGL J0634.6-3046c & 6h35m20.46499s ($\pm 0.83$) & -30d47m36.27438s ($\pm 0.56$) & 8.78 & -1.45 &  & T2 \\
J063522.62-304736.1 & 4FGL J0634.6-3046c & 6h35m22.62253s ($\pm 0.71$) & -30d47m36.161448s ($\pm 0.33$) & 9.24 & -2.13 &  & T2 \\
J195939.23-183924.2 & 4FGL J1959.8-1835 & 19h59m39.2393s ($\pm 0.77$) & -18d39m24.245748s ($\pm 0.61$) & 5.21 & -1.72 &  & T2 \\
J004136.39-010937.1 & 4FGL J0041.3-0048 & 0h41m36.38967s ($\pm 0.57$) & -1d09m37.1153412s ($\pm 0.71$) & 21.87 & -2.64 & 0.713 & RG \\
J013134.98-095325.9 & 4FGL J0132.1-0956 & 1h31m34.98756s ($\pm 0.36$) & -9d53m25.9342224s ($\pm 0.42$) & 8.47 & -2.67 &  & RG \\
J013234.52-100136.1 & 4FGL J0132.1-0956 & 1h32m34.52194s ($\pm 0.05$) & -10d01m36.17436s ($\pm 0.07$) & 8.53 & -1.44 & 0.579 & RG \\
J025142.87-270515.9 & 4FGL J0252.3-2707 & 2h51m42.87359s ($\pm 0.23$) & -27d05m15.91152s ($\pm 0.32$) & 8.71 & -2.5 & 1.05 & RG \\
J025157.96-265942.6 & 4FGL J0252.3-2707 & 2h51m57.96059s ($\pm 0.16$) & -26d59m42.609408s ($\pm 0.21$) & 9.53 & -1.41 & 1.052 & RG \\
J025218.21-271316.0 & 4FGL J0252.3-2707 & 2h52m18.2129s ($\pm 0.10$) & -27d13m16.058208s ($\pm 0.15$) & 5.43 & -1.45 & 0.436 & RG \\
J025243.39-271733.4 & 4FGL J0252.3-2707 & 2h52m43.39807s ($\pm 0.38$) & -27d17m33.478692s ($\pm 0.21$) & 10.99 & -2.0 & 0.165 & RG \\
J044325.75-594707.2 & 4FGL J0443.3-5949 & 4h43m25.75598s ($\pm 0.38$) & -59d47m07.2528s ($\pm 1.2$) & 2.72 & -1.71 &  & RG \\
J044335.09-594906.7 & 4FGL J0443.3-5949 & 4h43m35.09739s ($\pm 0.27$) & -59d49m06.743928s ($\pm 0.45$) & 2.1 &  -1.81 & 1.826 & RG \\
J060820.47-230127.7 & 4FGL J0608.6-2305 & 6h8m20.47386s ($\pm 0.61$) & -23d01m27.715296s ($\pm 1.05$) & 5.42 &  -1.8 &  & RG \\
J060841.66-230613.1 & 4FGL J0608.6-2305 & 6h8m41.66534s ($\pm 0.05$) & -23d06m13.133088s ($\pm 0.06$) & 1.46 & -1.45 &  & RG \\
J063347.19-304842.9 & 4FGL J0634.6-3046c & 6h33m47.19064s ($\pm 0.33$) & -30d48m42.986628s ($\pm 0.42$) & 11.62 & -1.57 &  & RG \\
J063347.20-304843.2 & 4FGL J0634.6-3046c & 6h33m47.20138s ($\pm 0.29$) & -30d48m43.212564s ($\pm 0.48$) & 11.62 & -1.43 &  & RG \\
J063408.80-305041.2 & 4FGL J0634.6-3046c & 6h34m8.80771s ($\pm 0.81$) & -30d50m41.225964s ($\pm 0.72$) & 8.05 & -1.86 &  & RG \\
J063424.94-303809.3 & 4FGL J0634.6-3046c & 6h34m24.944s ($\pm 0.34$) & -30d38m09.346308s ($\pm 0.25$) & 8.72 & -1.64 &  & RG \\
J063432.91-305803.8 & 4FGL J0634.6-3046c & 6h34m32.91927s ($\pm 0.84$) & -30d58m03.82152s ($\pm 0.55$) & 11.91 & -1.99 &  & RG \\
J063441.26-304701.5 & 4FGL J0634.6-3046c & 6h34m41.26876s ($\pm 0.31$) & -30d47m01.593744s ($\pm 0.23$) & 0.82 & -1.58 &  & RG \\
J063443.35-304338.8 & 4FGL J0634.6-3046c & 6h34m43.35245s ($\pm 0.75$) & -30d43m38.872596s ($\pm 1.02$) & 2.69 &  -3.0 &  & RG \\
J063506.41-305625.1 & 4FGL J0634.6-3046c & 6h35m6.41386s ($\pm 1.02$) & -30d56m25.133064s ($\pm 0.91$) & 11.64 &  -2.32 &  & RG \\
J063506.44-305624.8 & 4FGL J0634.6-3046c & 6h35m6.43996s ($\pm 1.36$) & -30d56m24.795096s ($\pm 0.72$) & 11.64 &  -1.76 &  & RG \\
J063518.93-304016.7 & 4FGL J0634.6-3046c & 6h35m18.9376s ($\pm 0.93$) & -30d40m16.75884s ($\pm 0.49$) & 10.27 & 2 -2.76 &  & RG \\
J063522.40-305337.6 & 4FGL J0634.6-3046c & 6h35m22.40418s ($\pm 0.35$) & -30d53m37.628556s ($\pm 0.30$) & 11.71 &  -1.73 &  & RG \\
J082100.90-064913.9 & 4FGL J0821.0-0651 & 8h21m0.906s ($\pm 0.13$) & -6d49m13.9931976s ($\pm 0.15$) & 1.92 &  -1.48 & 1.183 & RG \\
J094429.13-091337.8 & 4FGL J0944.3-0911 & 9h44m29.13362s ($\pm 0.44$) & -9d13m37.8679404s ($\pm 0.37$) & 2.79 &  -2.09 &  & RG \\
J132143.69-462635.7 & 4FGL J1322.0-4624 & 13h21m43.6963s ($\pm 0.51$) & -46d26m35.781288s ($\pm 0.72$) & 3.76 &  -1.43 &  & RG \\
J132155.24-462557.1 & 4FGL J1322.0-4624 & 13h21m55.24795s ($\pm 0.50$) & -46d25m57.174348s ($\pm 0.76$) & 1.73 &  -2.12 &  & RG \\
J133411.00-385504.9 & 4FGL J1334.8-3856 & 13h34m11.00909s ($\pm 0.88$) & -38d55m04.942164s ($\pm 1.06$) & 7.43 &  -4.45 &  & RG \\
J133436.33-390212.5 & 4FGL J1334.8-3856 & 13h34m36.33619s ($\pm 0.65$) & -39d02m12.5574s ($\pm 0.91$) & 6.21 &  -2.21 &  & RG \\
J133450.89-390354.6 & 4FGL J1334.8-3856 & 13h34m50.88953s ($\pm 1.13$) & -39d03m54.676404s ($\pm 0.76$) & 7.46 &  -1.54 &  & RG \\
J133506.65-385618.4 & 4FGL J1334.8-3856 & 13h35m6.65522s ($\pm 1.05$) & -38d56m18.39606s ($\pm 0.99$) & 3.53 &  -1.84 &  & RG \\
J133515.17-385258.5 & 4FGL J1334.8-3856 & 13h35m15.17621s ($\pm 0.46$) & -38d52m58.498824s ($\pm 0.40$)  & 6.25 &  -2.07 &  & RG \\
J133517.57-385834.7 & 4FGL J1334.8-3856 & 13h35m17.57114s ($\pm 1.44$) & -38d58m34.702896s ($\pm 1.39$) & 6.03 &  -2.33 &  & RG \\
J162101.76+002317.4 & 4FGL J1620.9+0021 & 16h21m1.76266s ($\pm 1.00$) & 0d23m17.4514344s ($\pm 1.41$) & 1.96 &  -1.97 & 0.451 & RG \\
J195908.94-184034.7 & 4FGL J1959.8-1835 & 19h59m8.94514s ($\pm 0.77$) & -18d40m34.781268s ($\pm 0.43$) & 11.63 &  -1.69 &  & RG \\
J195912.05-184237.5 & 4FGL J1959.8-1835 & 19h59m12.05333s ($\pm 0.11$) & -18d42m37.59408s ($\pm 0.08$) & 12.11 &  -1.61 &  & RG \\
J195920.98-184326.1 & 4FGL J1959.8-1835 & 19h59m20.98202s ($\pm 0.86$) & -18d43m26.183352s ($\pm 0.55$) & 11.09 &  -3.0 &  & RG \\
J195922.05-184741.6 & 4FGL J1959.8-1835 & 19h59m22.05454s ($\pm 1.00$) & -18d47m41.6436s ($\pm 0.64$) & 14.38 &  -1.47 &  & RG \\
J195941.56-184145.4 & 4FGL J1959.8-1835 & 19h59m41.56164s ($\pm 0.35$) & -18d41m45.40578s ($\pm 0.20$) & 7.0 &  -1.81 &  & RG \\
J195953.95-184035.9 & 4FGL J1959.8-1835 & 19h59m53.95891s ($\pm 1.28$) & -18d40m35.989104s ($\pm 0.49$) & 5.34 &  -1.64 &  & RG \\
J195955.80-184710.5 & 4FGL J1959.8-1835 & 19h59m55.80494s ($\pm 0.69$) & -18d47m10.50882s ($\pm 0.60$) & 11.93 &  -2.08 &  & RG \\
J203217.85-054325.2 & 4FGL J2032.2-0545 & 20h32m17.85053s ($\pm 0.18$) & -5d43m25.2195168s ($\pm 0.33$) & 2.13 &  -2.0 & 2.701 & RG
\enddata
\end{deluxetable}

\begin{appendix}

\begin{figure}[htb!]
\includegraphics[width=15cm]{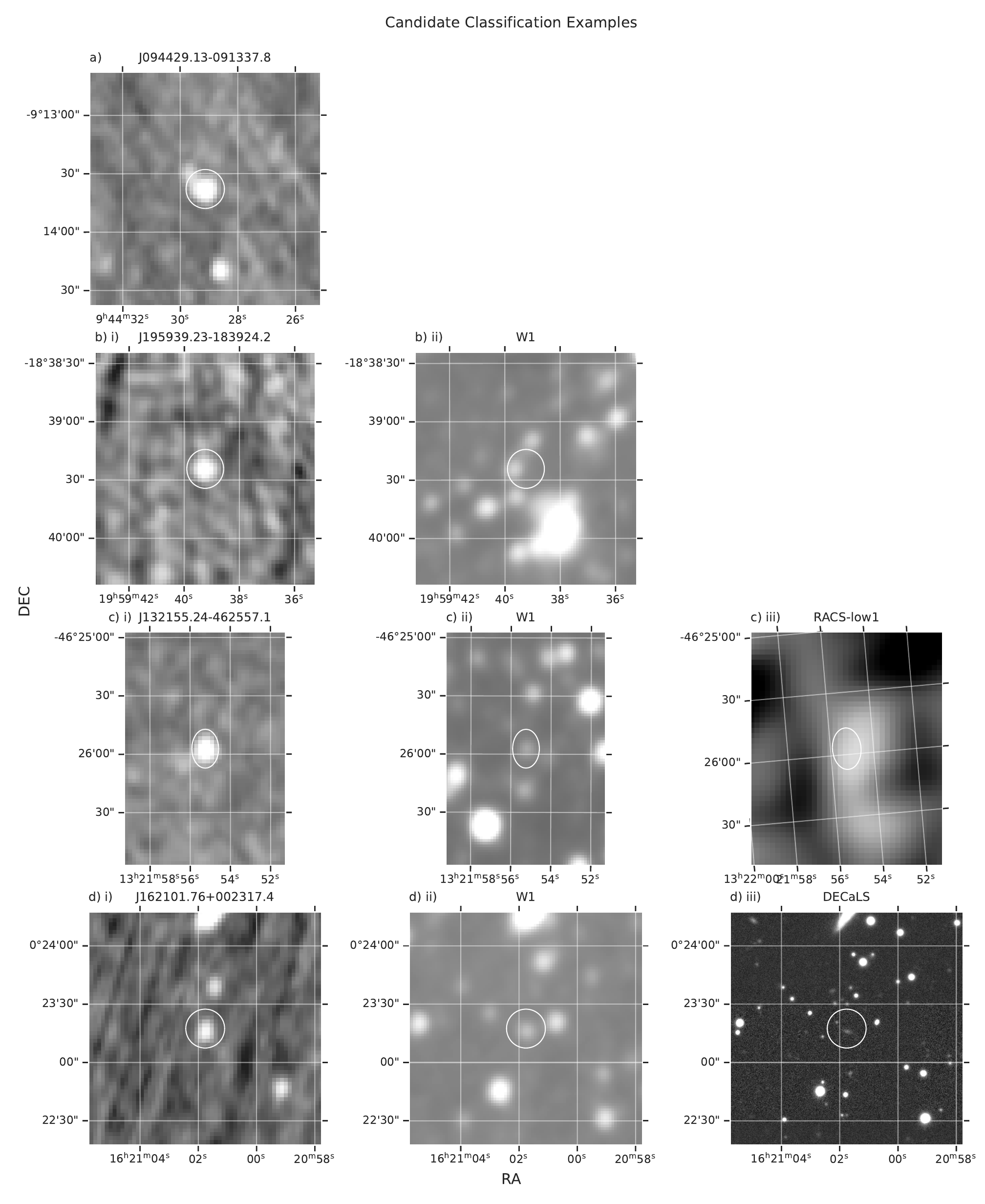}
\centering
\caption{Some examples of sources classified with the flowchart in Fig.~\ref{fig:flowchart} are shown. Subplot (a) shows an example of an extended source in MALS; though this source was marked as compact in the catalog, it appears extended in the wideband image. Subplot (b) is an example of a Tier 2 pulsar candidate with a WISE detection. The MALS detection is shown in (b)(i), and the WISE detection is shown in (b)(ii). Subplot (c) is an example of a source that is extended in RACS-low1. The MALS detection is shown in (c)(i), the WISE detection is shown in (c)(ii), and the RACS detection is shown in (c)(iii). Because RACS-low1 is lower resolution than MALS, we consider extended sources in RACS to be radio galaxy candidates. Subplot (d) is an example of a radio galaxy with detections in WISE and DECaLS. The MALS detection is shown in (d)(i), the WISE detection is shown in (d)(ii), and the DECaLS detection is shown in (d)(iii). The source in panel (d) was observed with the VLA; see Sec. \ref{sec:galcand} for details. Circles with a 10 arcsecond radius are drawn around the candidates.  \label{fig:classifications}}
\end{figure}

\begin{figure}[htb!]
\includegraphics[width=11cm]{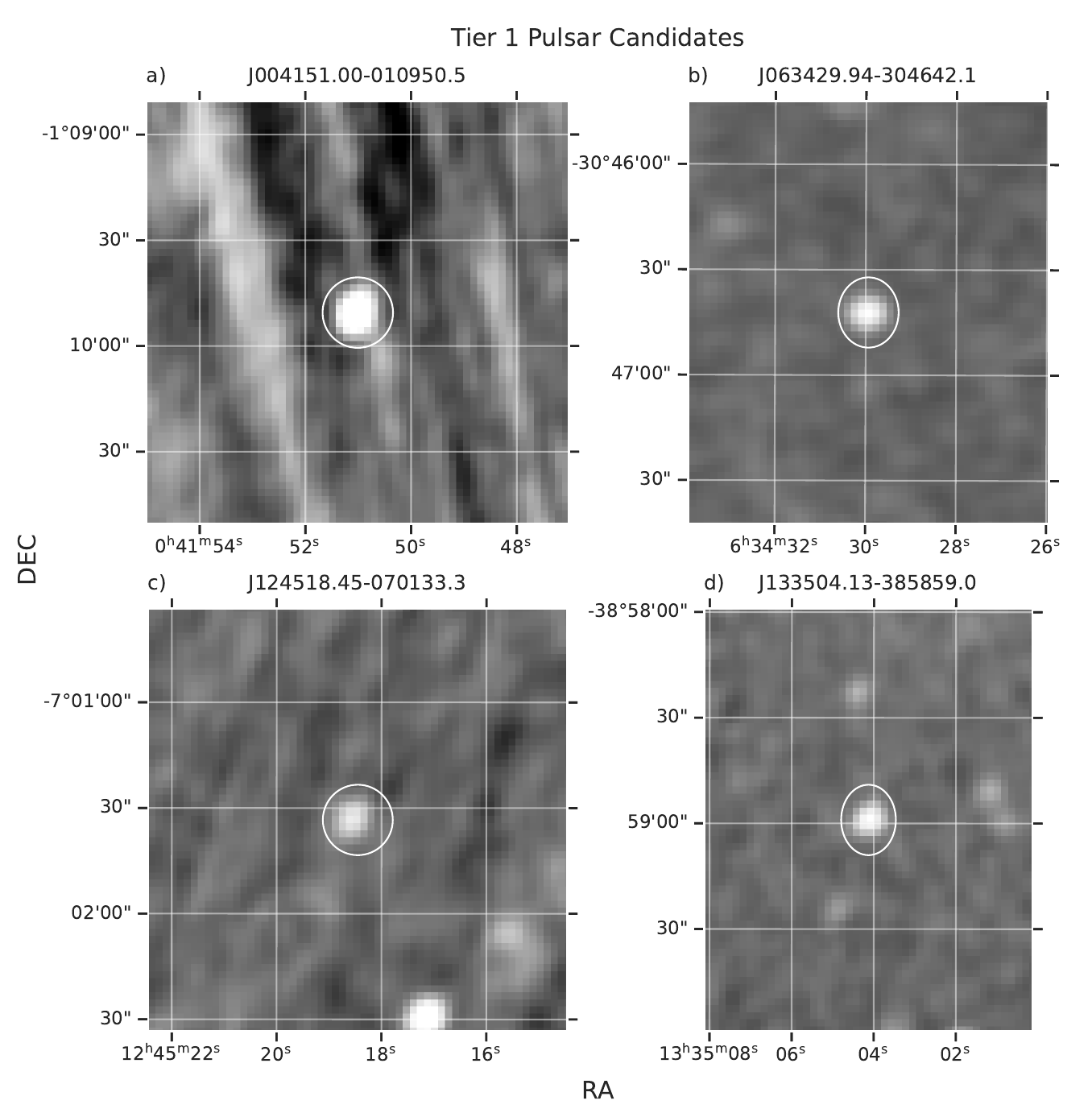}
\centering
\caption{Tier 1 pulsar candidates in MALS wideband images. Circles with a 10 arcsecond radius are drawn around the candidates. See Table~\ref{tab:tier1} for a description of each candidate. \label{fig:compactsources}}
\end{figure}

Examples of the sources classified based on our schema are included in Fig.~\ref{fig:classifications}. The four Tier 1 compact sources are shown in Fig.~\ref{fig:compactsources} for your reference.

\end{appendix}

\section{Acknowledgements} \label{sec:ack}
 We thank the referee for providing comments that improved the paper. This research has made use of the CIRADA cutout service at URL cutouts.cirada.ca, operated by the Canadian Initiative for Radio Astronomy Data Analysis (CIRADA). CIRADA is funded by a grant from the Canada Foundation for Innovation 2017 Innovation Fund (Project 35999), as well as by the Provinces of Ontario, British Columbia, Alberta, Manitoba and Quebec, in collaboration with the National Research Council of Canada, the US National Radio Astronomy Observatory and Australia’s Commonwealth Scientific and Industrial Research Organisation. 

This publication makes use of data products from the Wide-field Infrared Survey Explorer, which is a joint project of the University of California, Los Angeles, and the Jet Propulsion Laboratory/California Institute of Technology, funded by the National Aeronautics and Space Administration. This publication also makes use of data products from NEOWISE, which is a project of the Jet Propulsion Laboratory/California Institute of Technology, funded by the Planetary Science Division of the National Aeronautics and Space Administration. 

This scientific work uses data obtained from Inyarrimanha Ilgari Bundara / the Murchison Radio-astronomy Observatory. We acknowledge the Wajarri Yamaji People as the Traditional Owners and native title holders of the Observatory site. CSIRO’s ASKAP radio telescope is part of the Australia Telescope National Facility (https://ror.org/05qajvd42). Operation of ASKAP is funded by the Australian Government with support from the National Collaborative Research Infrastructure Strategy. ASKAP uses the resources of the Pawsey Supercomputing Research Centre. Establishment of ASKAP, Inyarrimanha Ilgari Bundara, the CSIRO Murchison Radio-astronomy Observatory and the Pawsey Supercomputing Research Centre are initiatives of the Australian Government, with support from the Government of Western Australia and the Science and Industry Endowment Fund. This paper includes archived data obtained through the CSIRO ASKAP Science Data Archive, CASDA (https://data.csiro.au).

The Pan-STARRS1 Surveys (PS1) and the PS1 public science archive have been made possible through contributions by the Institute for Astronomy, the University of Hawaii, the Pan-STARRS Project Office, the Max-Planck Society and its participating institutes, the Max Planck Institute for Astronomy, Heidelberg and the Max Planck Institute for Extraterrestrial Physics, Garching, The Johns Hopkins University, Durham University, the University of Edinburgh, the Queen's University Belfast, the Harvard-Smithsonian Center for Astrophysics, the Las Cumbres Observatory Global Telescope Network Incorporated, the National Central University of Taiwan, the Space Telescope Science Institute, the National Aeronautics and Space Administration under Grant No. NNX08AR22G issued through the Planetary Science Division of the NASA Science Mission Directorate, the National Science Foundation Grant No. AST-1238877, the University of Maryland, Eotvos Lorand University (ELTE), the Los Alamos National Laboratory, and the Gordon and Betty Moore Foundation.

The National Radio Astronomy Observatory is a facility of the National Science Foundation operated under cooperative agreement by Associated Universities, Inc.

The Legacy Surveys consist of three individual and complementary projects: the Dark Energy Camera Legacy Survey (DECaLS; Proposal ID \#2014B-0404; PIs: David Schlegel and Arjun Dey), the Beijing-Arizona Sky Survey (BASS; NOAO Prop. ID \#2015A-0801; PIs: Zhou Xu and Xiaohui Fan), and the Mayall z-band Legacy Survey (MzLS; Prop. ID \#2016A-0453; PI: Arjun Dey). DECaLS, BASS and MzLS together include data obtained, respectively, at the Blanco telescope, Cerro Tololo Inter-American Observatory, NSF’s NOIRLab; the Bok telescope, Steward Observatory, University of Arizona; and the Mayall telescope, Kitt Peak National Observatory, NOIRLab. Pipeline processing and analyses of the data were supported by NOIRLab and the Lawrence Berkeley National Laboratory (LBNL). The Legacy Surveys project is honored to be permitted to conduct astronomical research on Iolkam Du’ag (Kitt Peak), a mountain with particular significance to the Tohono O’odham Nation.

NOIRLab is operated by the Association of Universities for Research in Astronomy (AURA) under a cooperative agreement with the National Science Foundation. LBNL is managed by the Regents of the University of California under contract to the U.S. Department of Energy.

This project used data obtained with the Dark Energy Camera (DECam), which was constructed by the Dark Energy Survey (DES) collaboration. Funding for the DES Projects has been provided by the U.S. Department of Energy, the U.S. National Science Foundation, the Ministry of Science and Education of Spain, the Science and Technology Facilities Council of the United Kingdom, the Higher Education Funding Council for England, the National Center for Supercomputing Applications at the University of Illinois at Urbana-Champaign, the Kavli Institute of Cosmological Physics at the University of Chicago, Center for Cosmology and Astro-Particle Physics at the Ohio State University, the Mitchell Institute for Fundamental Physics and Astronomy at Texas A\&M University, Financiadora de Estudos e Projetos, Fundacao Carlos Chagas Filho de Amparo, Financiadora de Estudos e Projetos, Fundacao Carlos Chagas Filho de Amparo a Pesquisa do Estado do Rio de Janeiro, Conselho Nacional de Desenvolvimento Cientifico e Tecnologico and the Ministerio da Ciencia, Tecnologia e Inovacao, the Deutsche Forschungsgemeinschaft and the Collaborating Institutions in the Dark Energy Survey. The Collaborating Institutions are Argonne National Laboratory, the University of California at Santa Cruz, the University of Cambridge, Centro de Investigaciones Energeticas, Medioambientales y Tecnologicas-Madrid, the University of Chicago, University College London, the DES-Brazil Consortium, the University of Edinburgh, the Eidgenossische Technische Hochschule (ETH) Zurich, Fermi National Accelerator Laboratory, the University of Illinois at Urbana-Champaign, the Institut de Ciencies de l’Espai (IEEC/CSIC), the Institut de Fisica d’Altes Energies, Lawrence Berkeley National Laboratory, the Ludwig Maximilians Universitat Munchen and the associated Excellence Cluster Universe, the University of Michigan, NSF’s NOIRLab, the University of Nottingham, the Ohio State University, the University of Pennsylvania, the University of Portsmouth, SLAC National Accelerator Laboratory, Stanford University, the University of Sussex, and Texas A\&M University.

BASS is a key project of the Telescope Access Program (TAP), which has been funded by the National Astronomical Observatories of China, the Chinese Academy of Sciences (the Strategic Priority Research Program “The Emergence of Cosmological Structures” Grant \# XDB09000000), and the Special Fund for Astronomy from the Ministry of Finance. The BASS is also supported by the External Cooperation Program of Chinese Academy of Sciences (Grant \# 114A11KYSB20160057), and Chinese National Natural Science Foundation (Grant \# 12120101003, \# 11433005).

The Legacy Survey team makes use of data products from the Near-Earth Object Wide-field Infrared Survey Explorer (NEOWISE), which is a project of the Jet Propulsion Laboratory/California Institute of Technology. NEOWISE is funded by the National Aeronautics and Space Administration.

The Legacy Surveys imaging of the DESI footprint is supported by the Director, Office of Science, Office of High Energy Physics of the U.S. Department of Energy under Contract No. DE-AC02-05CH1123, by the National Energy Research Scientific Computing Center, a DOE Office of Science User Facility under the same contract; and by the U.S. National Science Foundation, Division of Astronomical Sciences under Contract No. AST-0950945 to NOAO.

The Photometric Redshifts for the Legacy Surveys (PRLS) catalog used in this paper was produced thanks to funding from the U.S. Department of Energy Office of Science, Office of High Energy Physics via grant DE-SC0007914.

This research makes use of the ATNF Pulsar Catalogue at URL atnf.csiro.au/research/pulsar/psrcat.

NG acknowledges NRAO for financial support for the sabbatical visit at Socorro during which (a part of) this work was done. 

%

\vspace{5mm}
\facilities{EVLA, MeerKAT, ASKAP}


\software{CASA \citep{CASA2007}, 
Astropy \citep{astropy:2013, astropy:2018}, Matplotlib \citep{Matplotlib}, Seaborn \citep{Seaborn}, TOPCAT \citep{taylor2005topcat}}




\bibliography{fermiuas}{}
\bibliographystyle{aasjournal}



\end{document}